\documentclass[11pt,a4paper]{article}

\usepackage[margin=1in]{geometry}

\usepackage{amsmath,amssymb,amsthm}
\usepackage{mathtools}

\usepackage{graphicx}
\usepackage{subcaption}
\usepackage{float}

\usepackage{setspace}

\usepackage{authblk}

\usepackage[
    colorlinks=true,
    linkcolor=blue,
    citecolor=blue,
    urlcolor=blue
]{hyperref}

\begin{document}

\singlespacing

\title{
    \textbf{
    Dynamical Tidal Response and Love Numbers of
    Massless Bosonic and Fermionic Perturbations
    of Kerr--Anti-de Sitter Black Holes
    }
}


\author[1]{
    Himanshu Buragohain
}

\author[1,2]{
    Prabwal Phukon
}

\affil[1]{
    Department of Physics, Dibrugarh University,
    Dibrugarh, Assam 786004, India
}

\affil[2]{
    Theoretical Physics Division, Centre for Atmospheric Studies,
    Dibrugarh University, Dibrugarh, Assam 786004, India
}

\affil[ ]{
    \texttt{rs\_himanshu1998buragohain@dibru.ac.in}
}

\affil[ ]{
    \texttt{prabwal@dibru.ac.in}
}

\date{}


\maketitle


\begin{abstract}
 We examine the tidal response
of Kerr--Anti-de Sitter black holes using the Teukolsky formalism in a vacuum background. We consider massless perturbations with spin weights
$
s = 0,\ \pm\frac{1}{2},\ \pm1,\ \pm\frac{3}{2},\ \pm2$, corresponding to scalar, fermionic, electromagnetic, and gravitational perturbations, respectively. We determine the Tidal Love number using the near-horizon approximation. For asymptotically flat black holes in general relativity, the conservative tidal response vanishes identically. In contrast, we find that the presence of a negative cosmological constant parameter in the
Kerr--Anti-de Sitter black holes leads to a non-trivial tidal response, containing both conservative and dissipative contributions. We systematically study the static and
dynamical tidal responses of Kerr--Anti-de Sitter black holes. In particular, we examine the behaviour of the conservative and dissipative response coefficients and investigate their dependence on the black-hole rotation parameter and the
AdS curvature scale.

\end{abstract}

	\maketitle
\section{Introduction}
In gravitational physics, tidal response coefficients provide a precise way to describe how compact objects react to external perturbations. In the relativistic regime, they naturally separate into a conservative part, which is encoded in the real part of the response, and a dissipative part, which is associated with the imaginary part of the response. When a star or planet is influenced by a heavy companion object, the companion's gravity pulls on it and changes its shape. At first, the idea of tidal deformability originated in classical geophysics due to the pioneering research of A.E.H. Love in the 1900s \cite{a}.\\

In basic physics, an external gravitational force creates a squeezing effect (a tidal field) around the body. The body responds by shifting its mass into a bulge (a quadrupole moment). The formula for this relation is given by:
\begin{equation}
    Q_{ab} = -\frac{2}{3} k_2 R^5 E_{ab},
\end{equation}
where, $Q_{ab}$ represents the induced trace-free quadrupole moment tensor and $E_{ab}$ is the external tidal field.Here, $R$ is the body's radius, and $k_2$ is the dimensionless Love number. This number acts as a window into the object's interior, telling us how its mass is packed together. While $k_2$ measures the simplest quadrupole bulge, we use higher-order numbers ($k_l$) for more complex multipole distortions.\\

In the realm of gravitational-wave astronomy, the extraction of TLNs serves as an essential diagnostic instrument to understand the internal equation of state of ultra-dense entities such as neutron stars\cite{n1,n2,n3,n4}, where the deformability of objects directly affects gravitational-wave signatures. In the era of precision gravitational-wave astronomy inaugurated by the significant observation of compact binary coalescences \cite{a1}, tidal deformability has emerged as a fundamental physical observable for investigating the internal composition and strong-field dynamics of compact objects. In the late inspiral phase of a compact binary merger, the gravitational field of each companion exerts a tidal force on the other, causing multipolar deformations that accelerate the orbital decay and leave a distinctive imprint on the emitted gravitational waveform \cite{n3,exp3}.\\

Previous studies in both Newtonian and relativistic frameworks have shown that a Schwarzschild black hole has zero tidal Love numbers \cite {n3,n4,stat1,stat4,stat5}. For a spinning black hole, dissipation can remain nonzero even for static perturbations \cite {im1,im2} due to frame dragging. Extending this picture to full general relativity requires a careful treatment of the two-body problem using post-Newtonian and matching techniques \cite{n2,b,89}. In this framework, the spacetime is divided into two regions. The outer region describes the binary system in the weak-field regime, where each compact object is represented by a worldline carrying its multipole moments \cite{eric2, mar, eric}. The inner region describes the spacetime close to each body, where the metric is expanded in terms of the external tidal field and the body's own multipole moments. Matching these two descriptions allows one to relate the induced multipole moments to the applied tidal fields, with the proportionality coefficients defining the relativistic Love numbers. These coefficients encode the compact object's finite-size response and enter the binary dynamics and gravitational-wave signal at high post-Newtonian order; for the leading quadrupolar response, the effect first appears at fifth post-Newtonian order. A closely related description arises in the point-particle effective field theory of compact binaries, where Love numbers are interpreted as Wilson coefficients of non-minimal worldline couplings to the Weyl tensor \cite{ef1,ef2,ef3}. This provides a convenient and gauge-invariant framework for separating the finite-size response of the compact objects from the point-particle dynamics.
When this framework is applied to black holes in general relativity, it leads to a remarkable result: the static tidal Love numbers of an isolated black hole vanish identically. This result was established through several independent approaches, including the matching of perturbed Schwarzschild solutions in both the Regge--Wheeler and light-cone gauges by Binnington and Poisson and by Damour and Nagar\cite{n3,n4}. The analysis was subsequently extended to rotating black holes. Studies by Le Tiec, Casals and Franzin\cite{im2}, as well as Landry and Poisson and Pani et al.\cite{p2}, showed that the conservative tidal response of a Kerr black hole also vanishes for axisymmetric perturbations, while non-axisymmetric perturbations can produce a non-zero response at higher orders in the spin. More recently, Charalambous, Dubovsky and Ivanov\cite{ch1} demonstrated that the vanishing of the static response persists for spin-$0$, spin-$1$, and spin-$2$ fields. They further related this property to a hidden $SL(2,\mathbb{R})$ symmetry of the near-zone Teukolsky equation, often referred to as the ``Love symmetry.\\

There is, however, something generic about this vanishing as long as you remain within the strict confines of four-dimensional general relativity. Whether the Love-number calculation is extended to higher spacetime dimensions, asymptotically non-flat or AdS backgrounds, black holes embedded in external distributions of matter, or even modified theories of gravity\cite{ch1,ch2,ch3,ch4,ch5,ch6,ch7,ch8,ch9,near1,ch11}, a common theme uncovered here has been the presence of non-vanishing tidal responses -- indicating that the conclusion reached in standard four-dimensional GR regarding vanishing tidal numbers reflects rather a particularly special and symmetry-protected structure than a universal property of horizons.\\

Most recently, this line of investigation has been pushed into two new directions relevant to the present work. First, it has been shown that the long-standing "vanishing" result is not universal even within four-dimensional GR once the spin of the perturbing field is changed: while bosonic (integer-spin) perturbations of a Kerr or Schwarzschild black hole give strictly zero static Love numbers, fermionic (half-integer-spin) perturbations do not. Chakraborty, Heidmann and Pani derived closed-form\cite{fm}, non-vanishing fermionic Love numbers for Dirac-type fields around a Kerr black hole of arbitrary spin, tracing this bosonic/fermionic asymmetry to a breaking of the same hidden symmetry responsible for the vanishing in the bosonic sector, while showing that the corresponding dissipative response still vanishes for static fermionic perturbations. Second, this bosonic-versus-fermionic distinction has been explored beyond black holes proper, in analogue-gravity systems: Du and Zhang computed both scalar (bosonic) and Dirac (fermionic) static Love numbers for acoustic black holes\cite{fm2}, finding a non-zero scalar response together with a universal power-law fermionic response, in contrast with the strict vanishing that characterises genuine general-relativistic black holes.\\

Our purpose in this work is to study the dynamical tidal response of rotating Kerr--AdS black holes to arbitrary spin-$s$ perturbations using the Teukolsky formalism \cite {bibi}. It consists of four sectors (electromagnetic, vector-fermionic, Rarita--Schwinger and gravitational) corresponding to  $s = \pm1, \pm\frac{1}{2}, \pm\frac{3}{2}, \pm2$. The static case of tidal response of static black holes has already been studied in \cite{ads}. We begin with the Teukolsky master equation in the Kerr--AdS background and, by separating radial and angular dependence, we obtain the corresponding near-zone radial solutions to obtain a frequency-dependent tidal response. Our calculation is based on the analytic near-zone solution and does not require a global matching procedure between the near-horizon and asymptotic AdS regions. It therefore provides a complementary approach to global matching methods, including those used in effective-field-theory descriptions, while allowing the frequency-dependent tidal response to be obtained directly from the near-zone dynamics.

\section{Kerr--AdS background and perturbation equations}
\label{sec:kerr_ads}

The perturbation equations for the Kerr geometry were originally formulated by Teukolsky \cite{teu1} within the Newman–Penrose formalism. These equations apply to Petrov type D backgrounds and therefore also describe perturbations of the Kerr–AdS spacetime. In the original formulation, Teukolsky employed an affinely parametrised null tetrad, commonly referred to as the outgoing Kinnersley tetrad, which remains regular across the past horizon \cite{teu2}.
We begin by considering linear perturbations of a four-dimensional
Kerr--AdS black hole. Our main objective is to establish the dynamical and static tidal response using radial
and angular perturbation equations, which have previously been studied in \cite{bibi,ads1,ads2}. The Kerr--AdS geometry is
characterised by the rotation parameter $a$, the mass parameter $M$,
and the AdS curvature radius $\ell_{\rm AdS}$.We have used the same parameters throughout the paper, which are already described in \cite{bibi}.  \\

The Kerr-AdS metric, originally discovered by Carter \cite{car}, is expressed in Boyer-Lindquist coordinates $\{\hat{t}, \hat{r}, \theta, \hat{\phi}\}$ as
$(\hat t,\hat r,\theta,\hat\phi)$, the Kerr--AdS metric can be written as
\begin{equation}
\begin{aligned}
ds^2={}&
-\frac{\Delta_r}{\Sigma^2}
\left(
d\hat t-\frac{a}{\Xi}\sin^2\theta\,d\hat\phi
\right)^2
+\frac{\Sigma^2}{\Delta_r}d\hat r^2
+\frac{\Sigma^2}{\Delta_\theta}d\theta^2
\\
&+
\frac{\Delta_\theta\sin^2\theta}{\Sigma^2}
\left(
a\,d\hat t
-\frac{\hat r^2+a^2}{\Xi}d\hat\phi
\right)^2 ,
\end{aligned}
\label{eq:kerr_ads_metric}
\end{equation}
where
\begin{equation}
\Delta_r
=
(\hat r^2+a^2)
\left(
1+\frac{\hat r^2}{\ell_{\rm AdS}^2}
\right)
-2M\hat r ,
\label{eq:delta_r}
\end{equation}
and
\begin{equation}
\Delta_\theta
=
1-\frac{a^2}{\ell_{\rm AdS}^2}\cos^2\theta,
\qquad
\Xi
=
1-\frac{a^2}{\ell_{\rm AdS}^2},
\end{equation}
with
\begin{equation}
\Sigma^2
=
\hat r^2+a^2\cos^2\theta.
\end{equation}

The event horizon is determined by the largest positive root of
\begin{equation}
\Delta_r(r_+)=0.
\label{eq:horizon_condition}
\end{equation}

The angular velocity of the horizon, measured relative to a non-rotating frame at infinity, is
\begin{equation}
\Omega_H
=
\frac{a}{r_+^2+a^2}
\left(
1+\frac{r_+^2}{\ell_{\rm Ads}^2}
\right).
\label{eq:omega_H}
\end{equation}

It is useful to emphasise that the AdS length introduces a new dimensionless parameter,
\begin{equation}
\frac{r_+}{\ell_{\rm AdS}},
\end{equation}
which controls the deviation from the asymptotically flat Kerr
geometry.\\
Since we are interested in the near-extremal and extremal configuration, we impose the conditions, which have already been described in \cite{bibi}
\begin{equation}
\Delta_r(r_+)=0,
\label{eq:extremal_conditions}
\end{equation}

Solving these conditions gives the extremal rotation parameter
\begin{equation}
a_{\rm ext}
=
r_+
\sqrt{
\frac{
3r_+^2+\ell_{\rm Ads}^2
}{
\ell_{\rm Ads}^2-r_+^2
}
}.
\label{eq:a_ext}
\end{equation}

The corresponding mass parameter is
\begin{equation}
{
M_{\rm ext}
=
\frac{
r_+
\left(
1+r_+^2/\ell_{\rm AdS}^2
\right)^2
}{
1-r_+^2/\ell_{\rm AdS}^2
}.
}
\label{eq:M_ext}
\end{equation}

Therefore, in the extremal limit, the angular velocity takes the form of
\begin{equation}
\Omega_H^{\rm ext}
=
\frac{
\sqrt{
\ell_{\rm AdS}^4
+2r_+^2\ell_{\rm AdS}^2
-3r_+^4
}
}{
2r_+\ell_{\rm AdS}^2
}.
\label{eq:omega_H_ext}
\end{equation}

The condition $a<\ell_{\rm AdS}$ further restricts the extremal solutions to
\begin{equation}
\boxed{
\frac{r_+}{\ell_{\rm AdS}}
<
\frac{1}{\sqrt{3}}.
}
\label{eq:extremal_bound}
\end{equation}
\\

The perturbations of the Kerr--AdS black hole can be described by the Teukolsky master equation. After separating the dependence on
the time and azimuthal coordinates, we write the master field as
\begin{equation}
\Psi^{(s)}(t,r,\theta,\phi)
=
R^{(s)}(r)\,
S^{(s)}(\theta)\,
e^{-i\omega t+im\phi},
\end{equation}
where $s$ denotes the spin weight, $\omega$ is the frequency,
$m$ is the azimuthal quantum number, and $R^{(s)}(r)$ and $S^{(s)}(\theta)$ are the radial and angular functions,respectively.\\

The radial part of the Teukolsky equation can be written as
\begin{equation}
\Delta_r^{-s}
\frac{d}{dr}
\left(
\Delta_r^{s+1}
\frac{dR^{(s)}}{dr}
\right)
+
H(r)R^{(s)}
=
0,
\label{eq:radial_teukolsky}
\end{equation}
where,

\begin{equation}
\begin{aligned}
H(\hat r)
={}&
\frac{K_T^2-is\Delta_{\hat r}'K_T}
{\Delta_{\hat r}}
+2isK_T'
+\frac{s+|s|}{2}\Delta_{\hat r}''
\\[2mm]
&-
\frac{|s|(|s|-1)(2|s|-1)(2|s|-7)}
{3\ell^2}\hat r^2
\\[2mm]
&-
\frac{|s|(|s|-2)
\left(4s^2-12|s|+11\right)}
{3\ell^2}a^2
-\hat{\lambda}^{(s)}_{\ell m\hat{\omega}} .
\end{aligned}
\end{equation}

Here, a prime denotes differentiation with respect to $r$, and the radial function $K_T(\hat r)$ is defined as,
\begin{equation}
K_T(\hat r)
=
\hat{\omega}(\hat r^2+a^2)
-
ma\left(
1+\frac{\hat r^2}{\ell^2}
\right).
\label{eq:KT}
\end{equation}

The radial function $\Delta_r$ appearing in
Eq.~\eqref{eq:radial_teukolsky} is
\begin{equation}
\Delta_r
=
(r^2+a^2)
\left(
1+\frac{r^2}{\ell_{\rm AdS}^2}
\right)
-2Mr .
\label{eq:Delta_r}
\end{equation}

while the angular sector obeys the Kerr--AdS spin-weighted spheroidal equation
\begin{equation}
\frac{1}{\sin\theta}(\sin\theta\,\Delta_\theta S')'
+\left[\frac{(a\omega\cos\theta)^2\Xi-2sa\omega\cos\theta\,\Xi}{\Delta_\theta}
+s+\Lambda^{(s)}_{\ell m\omega}
-\frac{\Delta_\theta}{\sin^2\theta}
\left(m+\frac{s\Xi\cos\theta}{\Delta_\theta}\right)^2
-\frac{2\delta_s a^2}{\ell_{\rm AdS}^2}\sin^2\theta\right]S=0,
\end{equation}
with
\begin{equation}
\lambda^{(s)}_{\ell m\omega}
=\Lambda^{(s)}_{\ell m\omega}-2am\omega+a^2\omega^2+(s+|s|).
\end{equation}

This frequency dependence is important for the dynamical tidal response because the angular sector cannot, in general, be replaced
by a frequency-independent spherical-harmonic eigenvalue.\\

In our case, we study the small-Ads deformation limit. 
In the small-rotation limit
\begin{equation}
\frac{a}{\ell_{\rm AdS}}\ll1,
\end{equation}
the Kerr--AdS angular eigenvalue can be approximated at leading order by
\begin{equation}
\boxed{
\Lambda^{(s)}_{\ell m\omega}
=
(\ell-s)(\ell+s+1)
+
\mathcal{O}\!\left(\frac{a}{\ell_{\rm AdS}}\right).
}
\label{eq:Lambda_small_a}
\end{equation}

Thus, the Kerr--AdS angular eigenvalue continuously approaches the spin-weighted spherical-harmonic eigenvalue as the AdS deformation becomes small.

\section{Calculation of the tidal response}
\label{sec:tidal_response}

In this section, we investigate the static and dynamical tidal response of a Kerr--AdS black hole within the near-zone approximation. We begin with the full Kerr--AdS Teukolsky radial equation and retain the essential near-horizon structure of the radial potential, which has been used in Ref\cite{near1,near2}. By expanding the radial function about the outer horizon and introducing a suitable dimensionless radial coordinate, the resulting equation can be transformed into a hypergeometric differential equation. This allows us to impose the physically relevant ingoing boundary condition at the horizon and analytically continue the solution to the overlap region, where the two independent radial branches can be identified as the tidal source and induced response. The ratio of these two coefficients then provides the frequency-dependent near-zone tidal response for a general Teukolsky spin $s$. The static response follows directly from the $\omega\rightarrow0$ limit of the dynamical result. Consequently, the quantity obtained here is a near-zone response coefficient; a globally normalised Kerr--AdS Love number would require an additional matching of the near-zone solution to the far-zone AdS solution.\\
To obtain an analytic description of the tidal response, we focus on the region immediately outside the outer event horizon, $r\simeq r_+$. In this regime, the radial function $\Delta_r$ varies rapidly near its zero at
$r=r_+$, while the remaining slowly varying factors in the radial potential can be evaluated at the horizon. All this information is described in Ref.\cite{near2}.We therefore expand $\Delta_r$ about the outer horizon as
\begin{equation}
\Delta_r(r)
\simeq
\Delta_r(r_+)
+(r-r_+)\Delta_r'(r_+)
+\frac{1}{2}(r-r_+)^2\Delta_r''(r_+).
\end{equation}
Since $\Delta_r(r_+)=0$, this can be written in the convenient factorized form
\begin{equation}
{
\Delta_r(r)
\simeq
k_{\rm Kerr}(r-r_+)(r-r_*)
}.
\end{equation}
The coefficient $k_{\rm Kerr}$ and the auxiliary root $r_*$ are determined
by matching the first and second derivatives at $r=r_+$:
\begin{equation}
{
k_{\rm Kerr}
=
\frac{1}{2}\Delta_r''(r_+)
},
\qquad
{
r_*
=
r_+
-\frac{\Delta_r'(r_+)}{k_{\rm Kerr}}
}.
\end{equation}

For the Kerr--AdS radial function
\begin{equation}
\Delta_r
=
(r^2+a^2)
\left(1+\frac{r^2}{\ell_{\rm AdS}^2}\right)
-2Mr,
\end{equation}
These quantities take the explicit form
\begin{equation}
{
k_{\rm Kerr}
=
1+\frac{a^2}{\ell_{\rm AdS}^2}
+\frac{6r_+^2}{\ell_{\rm AdS}^2}
},
\end{equation}
and
\begin{equation}
{
r_*
=
r_+
-\frac{1}{k_{\rm Kerr}r_+}
\left[
r_+^2-a^2
+\frac{3r_+^4}{\ell_{\rm AdS}^2}
+\frac{a^2r_+^2}{\ell_{\rm AdS}^2}
\right].
}
\end{equation}
Here $r_*$ is an auxiliary root associated with the quadratic near-horizon approximation and should not be identified with an additional
physical horizon. This approximation preserves the leading radial structure of $\Delta_r$ in the vicinity of $r_+$. 

The radial Teukolsky equation can be written in the form of
\begin{equation}
\Delta_r R''+
(s+1)\Delta_r'R'
+H(r)R=0 .
\label{eq:radial_compact}
\end{equation}

In the near-horizon region, the Kerr--AdS radial function is approximated by
\begin{equation}
\Delta_r
\simeq
k(r-r_+)(r-r_*),
\qquad
k=\frac{1}{2}\Delta_r''(r_+),
\label{eq:delta_near}
\end{equation}
where
\begin{equation}
d\equiv r_+-r_*
=
\frac{\Delta_r'(r_+)}{k}.
\label{eq:d_definition}
\end{equation}

We introduce the dimensionless near-zone coordinate
\begin{equation}
z=\frac{r-r_+}{r-r_*},
\label{eq:z_coordinate}
\end{equation}
for which
\begin{equation}
r-r_+
=
\frac{dz}{1-z},
\qquad
r-r_*=
\frac{d}{1-z},
\end{equation}
and consequently
\begin{equation}
\Delta_r
=
kd^2\frac{z}{(1-z)^2}.
\end{equation}

The horizon value of the radial momentum is
\begin{equation}
K_+
\equiv K_T(r_+)
=
(r_+^2+a^2)
(\omega-m\Omega_H),
\label{eq:Kplus}
\end{equation}
which motivates the dimensionless horizon-frequency parameter
\begin{equation}
{
\chi
\equiv
\frac{K_+}{\Delta_r'(r_+)}
=
\frac{(r_+^2+a^2)(\omega-m\Omega_H)}
{\Delta_r'(r_+)} .
}
\label{eq:chi}
\end{equation}

The remaining frequency- and spin-dependent terms can be collected
into
\begin{equation}
{
U_s=
\frac{
2isK'_+
+(s+|s|)k
-C_s^+
-\lambda^{(s)}_{\ell m\omega}
}{k},
}
\label{eq:Us}
\end{equation}
where
\begin{equation}
\lambda^{(s)}_{\ell m\omega}
=
\Lambda^{(s)}_{\ell m\omega}
-2am\omega
+a^2\omega^2
+(s+|s|)
\label{eq:lambda_response}
\end{equation}
and $C_s^+$ contains the curvature-dependent Kerr--AdS terms evaluated
at the horizon.

With these definitions, the near-zone radial equation takes the form
\begin{equation}
{
z(1-z)R''
+
\left[s+1+(s-1)z\right]R'
+
\left[
\frac{\chi(\chi-is)}{z}
-\chi(\chi+is)
+\frac{U_s}{1-z}
\right]R=0 .
}
\label{eq:near_zone_master}
\end{equation}

This equation has three regular singular points at
$z=0$, $z=1$, and $z=\infty$, and can therefore be transformed into
the Gauss hypergeometric equation.

We introduce
\begin{equation}
R(z)
=
z^{-s-i\chi}
(1-z)^{s-L_s}F(z),
\label{eq:R_hypergeom_ansatz}
\end{equation}
where the effective radial index is defined by
\begin{equation}
{
L_s
=
\frac{
\sqrt{(2s+1)^2-4U_s}-1
}{2}.
}
\label{eq:L_s}
\end{equation}

The function $F(z)$ then satisfies
\begin{equation}
z(1-z)F''
+
\left[
c-(a_h+b_h+1)z
\right]F'
-a_hb_hF=0,
\label{eq:hypergeom_standard}
\end{equation}
with the hypergeometric parameters
\begin{equation}
{
a_h=-L_s-s,
}
\label{eq:ah}
\end{equation}
\begin{equation}
{
b_h=-L_s-2i\chi,
}
\label{eq:bh}
\end{equation}
and
\begin{equation}
{
c=1-s-2i\chi.
}
\label{eq:ch}
\end{equation}

The solution satisfying the ingoing boundary condition at the future
horizon is therefore
\begin{equation}
{
R_{\rm in}(z)
=
z^{-s-i\chi}
(1-z)^{s-L_s}
\,{}_2F_1
\left(
-L_s-s,\,
-L_s-2i\chi;\,
1-s-2i\chi;\,
z
\right).
}
\label{eq:ingoing_solution}
\end{equation}

The exponent $-s-i\chi$ is selected by the future-horizon ingoing condition. The two independent large-$r$ behaviours generated after analytic continuation towards $z\rightarrow1$ are
\begin{equation}
R_{\rm in}(r)
\sim
A_{\rm src}\,r^{L_s-s}
+
A_{\rm resp}\,r^{-L_s-s-1}.
\label{eq:source_response_asymptotic}
\end{equation}

The first term represents the applied tidal field, while the second term represents the induced response of the black hole.

We emphasise that the approximation does not constitute a solution of the full global Kerr--AdS boundary-value problem. In particular,
the AdS boundary condition cannot be imposed simply by extending the near-zone solution to arbitrarily large $r$. The quantity obtained below should consequently be interpreted as a near-zone dynamical response coefficient until a far-zone matching prescription and a normalisation of the tidal field have been specified.

\subsection{General dynamical tidal response}

Using the hypergeometric connection formula in the limit
$z\rightarrow1$, the two independent radial behaviours can be written
as
\begin{equation}
R_{\rm in}(r)
\sim
A_{\rm src}\,r^{L_s-s}
+
A_{\rm resp}\,r^{-L_s-s-1}.
\end{equation}

The dynamical response is defined by the ratio
\begin{equation}
\mathcal{R}^{(s)}_{\ell m}(\omega)
\equiv
\frac{A_{\rm resp}}{A_{\rm src}}.
\label{eq:R_response_definition}
\end{equation}

After cancellation of the common gamma-function factor, this gives
\begin{equation}
\boxed{
\mathcal{R}^{(s)}_{\ell m}(\omega)
=
d^{\,2L_s+1}
\frac{
\Gamma(-2L_s-1)
\Gamma(L_s+1-s)
\Gamma(L_s+1-2i\chi)
}{
\Gamma(2L_s+1)
\Gamma(-L_s-s)
\Gamma(-L_s-2i\chi)
}.
}
\label{eq:general_response}
\end{equation}

It is convenient to remove the dimensional factor
$d^{\,2L_s+1}$ and introduce the dimensionless response coefficient
\begin{equation}
\boxed{
\kappa^{(s)}_{\ell m}(\omega)
=
\frac{\mathcal{R}^{(s)}_{\ell m}(\omega)}
{d^{\,2L_s+1}} .
}
\label{eq:kappa_definition}
\end{equation}

Hence
\begin{equation}
\boxed{
\kappa^{(s)}_{\ell m}(\omega)
=
\frac{
\Gamma(-2L_s-1)
\Gamma(L_s+1-s)
\Gamma(L_s+1-2i\chi)
}{
\Gamma(2L_s+1)
\Gamma(-L_s-s)
\Gamma(-L_s-2i\chi)
}.
}
\label{eq:kappa_general}
\end{equation}

The dynamical response can be decomposed into conservative and
dissipative contributions,
\begin{equation}
\boxed{
\kappa^{(s)}_{\ell m}(\omega)
=
\kappa^{(s),R}_{\ell m}(\omega)
+
i\kappa^{(s),I}_{\ell m}(\omega),
}
\label{eq:kappa_decomposition}
\end{equation}
with
\begin{equation}
\kappa^{(s),R}_{\ell m}
=
\operatorname{Re}
\left[
\kappa^{(s)}_{\ell m}
\right],
\qquad
\kappa^{(s),I}_{\ell m}
=
\operatorname{Im}
\left[
\kappa^{(s)}_{\ell m}
\right].
\end{equation}

The real part describes the conservative deformation of the black hole induced by the external tidal field, whereas the imaginary part captures the dissipative component associated with the absorptive horizon response. In a rotating geometry, both components depend on the horizon-frame frequency through
\begin{equation}
{
\chi
=
\frac{(r_+^2+a^2)}
{\Delta_r'(r_+)}
\left(
\omega-m\Omega_H
\right).
}
\label{eq:chi_final}
\end{equation}

Thus, the relevant frequency variable for the horizon dynamics is not
$\omega$ alone, but
\begin{equation}
\omega_H=\omega-m\Omega_H.
\label{eq:horizon_frequency}
\end{equation}

The dimensionless dynamical tidal response obtained from the near-zone
radial solution can be expressed, for a general spin $s$, as
\begin{equation}
\boxed{
\kappa_{\ell m}^{(s)}(\omega)
=
\frac{
\sin[\pi(L_s+s)]
\sin[\pi(L_s+2i\chi)]
}{
\pi(2L_s+1)\sin(2\pi L_s)
}
\frac{
\Gamma(L_s+1-s)\Gamma(L_s+1+s)
\Gamma(L_s+1-2i\chi)
\Gamma(L_s+1+2i\chi)
}{
[\Gamma(2L_s+1)]^2
}.
}
\label{eq:general_dynamical_response}
\end{equation}

The curvature contribution entering $U_s$ is
\begin{equation}
{
\begin{aligned}
C_s^+
={}&
\frac{|s|(|s|-1)(2|s|-1)(2|s|-7)}
{3\ell_{\rm AdS}^{2}}\,r_+^2
\\
&+
\frac{|s|(|s|-2)(4s^2-12|s|+11)}
{3\ell_{\rm AdS}^{2}}\,a^2 ,
\end{aligned}
}
\label{eq:Cs_general}
\end{equation}

For real $\omega$, the Gamma-function pair satisfies
\begin{equation}
\Gamma(L_s+1-2i\chi)
\Gamma(L_s+1+2i\chi)
=
\left|
\Gamma(L_s+1+2i\chi)
\right|^2 ,
\end{equation}
so that all explicit complex dependence in Eq.~\eqref{eq:general_dynamical_response}
is contained in
\begin{equation}
\sin[\pi(L_s+2i\chi)].
\end{equation}
Using
\begin{equation}
\sin[\pi(L_s+2i\chi)]
=
\sin(\pi L_s)\cosh(2\pi\chi)
+i\cos(\pi L_s)\sinh(2\pi\chi),
\end{equation}
we write
\begin{equation}
\boxed{
\kappa_{\ell m}^{(s)}(\omega)
=
\operatorname{Re}\kappa_{\ell m}^{(s)}(\omega)
+i\,\operatorname{Im}\kappa_{\ell m}^{(s)}(\omega).
}
\label{eq:kappa_real_imag_general}
\end{equation}

Defining
\begin{equation}
\mathcal{P}_s(L_s,\chi)
=
\frac{
\sin[\pi(L_s+s)]
}{
\pi(2L_s+1)\sin(2\pi L_s)
}
\frac{
\Gamma(L_s+1-s)\Gamma(L_s+1+s)
\left|\Gamma(L_s+1+2i\chi)\right|^2
}{
[\Gamma(2L_s+1)]^2
},
\label{eq:P_general}
\end{equation}
The conservative and dissipative parts are explicitly
\begin{equation}
\boxed{
\operatorname{Re}\kappa_{\ell m}^{(s)}(\omega)
=
\mathcal{P}_s(L_s,\chi)
\sin(\pi L_s)
\cosh(2\pi\chi),
}
\label{eq:real_response_general}
\end{equation}
and
\begin{equation}
\boxed{
\operatorname{Im}\kappa_{\ell m}^{(s)}(\omega)
=
\mathcal{P}_s(L_s,\chi)
\cos(\pi L_s)
\sinh(2\pi\chi).
}
\label{eq:imag_response_general}
\end{equation}
 
For dynamical cases, we see both real and imaginary parts for the tidal response. The real part describes the conservative nature, and in the case of imaginary parts, which reflects about dissipation parts which is evolved due to the frame-dragging contribution \cite{stat1,im2} of the Love number and are directly related to the absorptive and emissive characteristics of the black hole horizon \cite{abs1,abs2}.\\
At this point, we want to mention that for all analytical and mathematical derivations throughout this work, we use the conventions and framework established in Ref.\cite{m1,m2}, and we also rely on MATHEMATICA and PYTHON for computation of numerical evaluations and the generation of plots presented in this work.
\subsection{Tidal response for Static limit}
In this section, we investigate the static case of Tidal response for Kerr AdS. To get the static Tidal response, we consider a stationary tidal perturbation by taking the static limit,
\begin{equation}
{\omega=0},
\end{equation}
and keeping the black-hole rotation parameter $a$ finite. In this limit, our parameters become
\begin{equation}
K_+=-ma\left(1+\frac{r_+^2}{\ell_{\rm AdS}^2}\right),
\qquad
K_+'=-\frac{2mar_+}{\ell_{\rm AdS}^2},
\end{equation}
and the horizon parameter takes the form of
\begin{equation}
{
\chi_0
=
-\frac{ma\left(1+r_+^2/\ell_{\rm AdS}^2\right)}
{\Delta_r'(r_+)}
}.
\end{equation}

Therefore, the dimensionless static near-zone tidal response has the form of
\begin{equation}
  \boxed{ {
\kappa_{\ell m}^{(s)}
=
\frac{
\Gamma(-2L_s-1)\,
\Gamma(L_s+1-s)\,
\Gamma(L_s+1-2i\chi_0)
}{
\Gamma(2L_s+1)\,
\Gamma(-L_s-s)\,
\Gamma(-L_s-2i\chi_0)
}
}}.
\end{equation}

Where,
\begin{equation}
{
L_s^{\rm stat}
=
\frac{1}{2}
\left[
\sqrt{(2s+1)^2-4U_s^{\rm stat}}-1
\right].
}
\end{equation}
and

\begin{equation}
{
U_s^{\rm stat}
=
\frac{
2isK_+'+(s+|s|)k-C_s^+
-\lambda_{\ell m0}^{(s)}
}{k}
}
\end{equation}

Here,
\begin{equation}
\lambda_{\ell m0}^{(s)}
=
\Lambda_{\ell m0}^{(s)}+s+|s|,
\end{equation}
and $C_s^+$ denotes the Kerr--AdS curvature contribution evaluated at the horizon.
\section{Discussions and Conclusions}
\begin{figure}[htbp]
    \centering

    \begin{subfigure}[b]{0.47\textwidth}
        \centering
        \includegraphics[width=\textwidth]{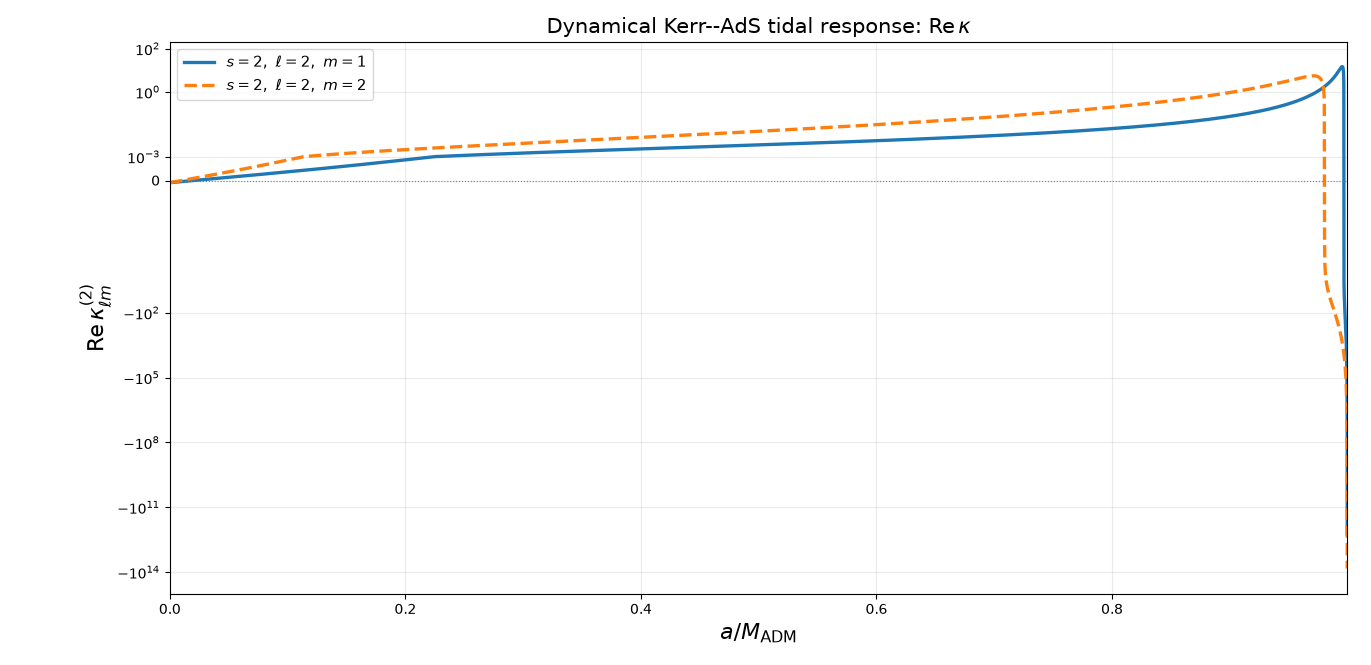}
        \caption{Real part for $s=2$.}
        \label{fig:s2_real}
    \end{subfigure}
    \hfill
    \begin{subfigure}[b]{0.47\textwidth}
        \centering
        \includegraphics[width=\textwidth]{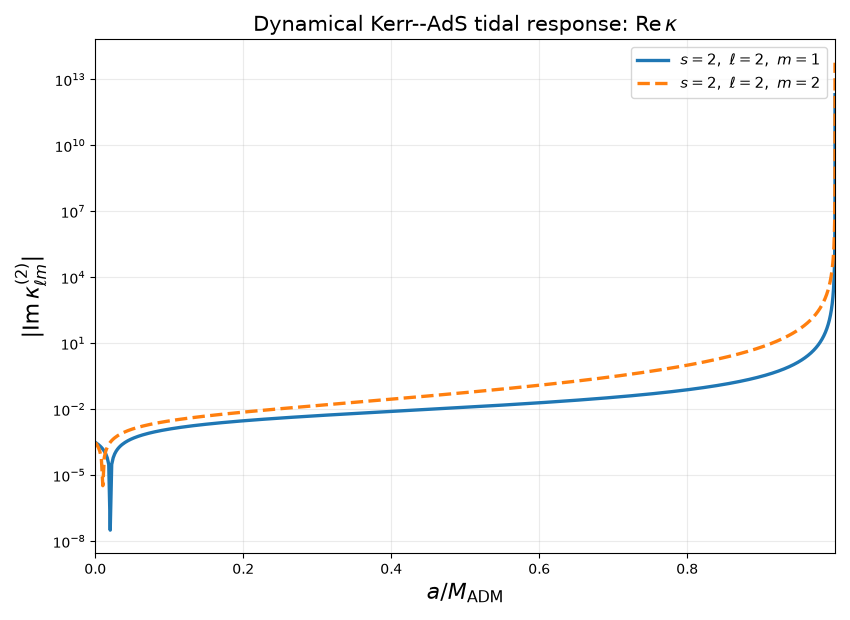}
        \caption{Imaginary part for $s=2$.}
        \label{fig:s2_imag}
    \end{subfigure}

    \vspace{0.4cm}

    \begin{subfigure}[b]{0.47\textwidth}
        \centering
        \includegraphics[width=\textwidth]{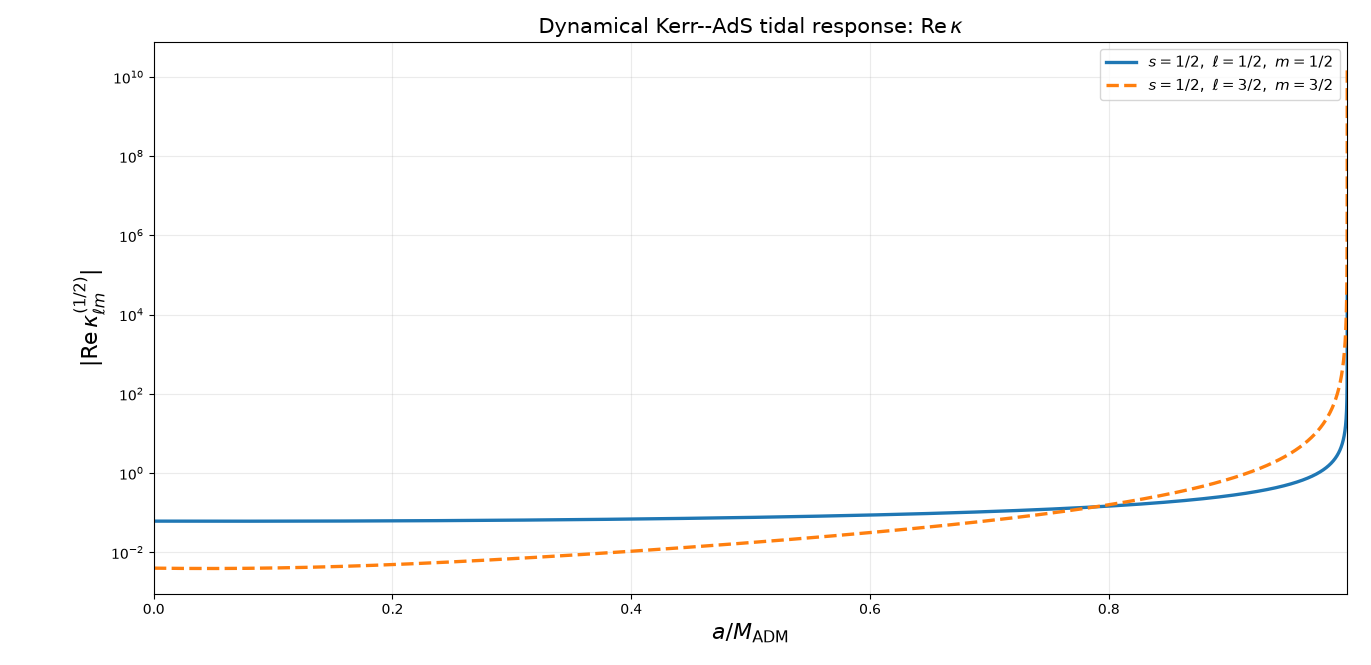}
        \caption{Real part for $s=\frac{1}{2}$.}
        \label{fig:s12_real}
    \end{subfigure}
    \hfill
    \begin{subfigure}[b]{0.47\textwidth}
        \centering
        \includegraphics[width=\textwidth]{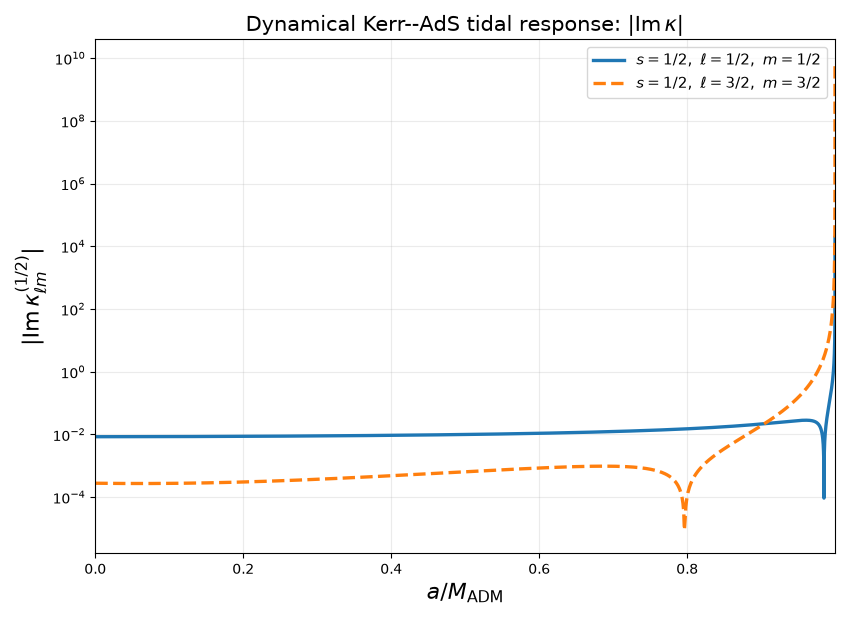}
        \caption{Imaginary part for $s=\frac{1}{2}$.}
        \label{fig:s12_imag}
    \end{subfigure}

    \vspace{0.4cm}

    \begin{subfigure}[b]{0.47\textwidth}
        \centering
        \includegraphics[width=\textwidth]{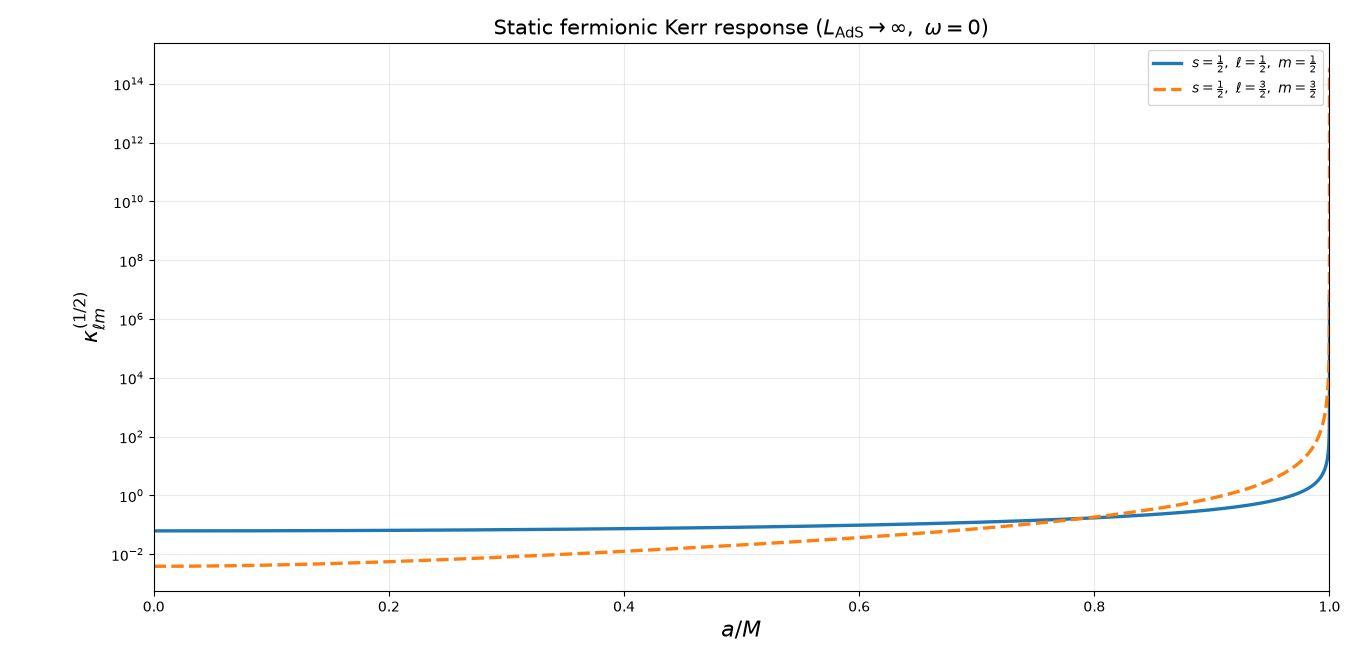}
        \caption{Static flat-space limit for $s=\frac{1}{2}$.}
        \label{fig:s12_static_flat}
    \end{subfigure}

    \caption{
        Tidal response of the Kerr--AdS black hole as a function of
        the dimensionless angular momentum $a/M_{\rm ADM}$. The first
        two plots show the real and imaginary parts of the response
        for gravitational perturbations with $s=2$, while the next
        two plots show the corresponding real and imaginary parts
        for fermionic perturbations with $s=1/2$. The final plot
        displays the fermionic response in the static flat-space
        limit. The comparison illustrates the dependence of the
        tidal response on the angular momentum and on the spin of
        the perturbing field. In the static flat-space limit, the
        fermionic response becomes purely real and exhibits a
        qualitative rotational dependence similar to the fermionic
        Love-number behaviour previously obtained for Kerr black
        holes \cite{fm}.
    }
    \label{fig:kerrads_tidal_response}
\end{figure}

 We have studied an analytical framework to determine both dynamical and static tidal responses for a Kerr Anti-de Sitter black hole. To obtain our numerical results for the tidal response, we employed the near-horizon approximation. In particular, we find that the presence of a finite Anti-de Sitter (AdS) curvature radius/radius of curvature introduces both conservative (real) and dissipative (imaginary) components to the tidal response function in the Kerr--AdS system. However, in the asymptotically flat Kerr limit, the conservative tidal response vanishes for bosonic ($s=2$) perturbations, while the dissipative component vanishes for fermionic ($s=1/2$) perturbations.\\ 

Interestingly, the imaginary component of the response function is non-zero for both $s=2$ and $s=1/2$ spin contributions, reflecting the intrinsic absorptive and emissive nature of the black hole horizon~\cite{30,31,32,33,34,35}. Consequently, this imaginary contribution acts as a macroscopic measure of dissipation and is directly proportional to the net energy flux crossing the horizon.\\

Based on the mathematical findings, we subsequently investigate the dependence of the tidal response function on the angular momentum parameter. The five plots in Fig.\ref{fig:kerrads_tidal_response} provide a systematic comparison of the tidal response for varying angular momentum parameter, revealing the behaviour for gravitational ($s=2$) and fermionic ($s=1/2$) perturbations. Both the conservative and dissipative responses in the dynamical case show a nontrivial dependence on the Kerr--AdS black hole's rotation when we plot them as functions of the dimensionless angular momentum, $a/M_{\rm ADM}$. The $s=2$ and $s=1/2$ sectors show qualitatively different rotational behaviour, reflecting how the relevant perturbation equations depend on spin and their modes.\\

Interestingly, the fermionic response sector is particularly interesting because its rotational dependence persists in the static limit. As a consistency check, we consider the static flat-space limit, obtained by taking
$\omega\rightarrow0$ and $\ell_{\rm AdS}\rightarrow\infty$. In the static flat-space limit, $\omega\rightarrow0$ and
$\ell_{\rm AdS}\rightarrow\infty$, the imaginary part of the fermionic response vanishes; Fig \ref{fig:kerrads_tidal_response}(e) display about kerr flat limit for the static response, and the response becomes purely real, which is consistent with the results of Ref.~\cite{fm}, where they studied the static fermionic response of the Kerr black hole. The main difference is the presence of a finite AdS curvature scale ($\ell_{\rm AdS}$) and a finite frequency, which gives rise to a dissipative imaginary piece associated with the dynamical horizon response. Thus, the Kerr--AdS system retains the intrinsic real fermionic tidal response of Kerr, but also acquires a novel dynamical imaginary part. The primary difference is the presence of a finite AdS curvature scale $\ell_{\rm AdS}$ and a finite frequency, which gives rise to a dissipative imaginary contribution associated with the dynamical horizon response. As a result, the Kerr--AdS system retains the intrinsic real fermionic static tidal response of Kerr in flat space, but also develops a non-vanishing imaginary component, which denotes the effects of dynamical response and AdS curvature parameters.\\

Remarkably, for the static, nonrotating and asymptotically flat limit, the Kerr--AdS response coefficient reduces to the Schwarzschild result. For the fermionic perturbation with $s=1/2$, we derive a non-trivial tidal response in the asymptotically flat Kerr limit, which takes the form of
\begin{equation}
{
\kappa_{\ell}^{(\pm 1/2)}
=
\pm\,4^{-2\ell-1},
\qquad
\ell=\frac12,\frac32,\frac52,\ldots .
}
\end{equation}
This is in perfect agreement with the known Schwarzschild fermionic response found in the corresponding Kerr analysis \cite{fm},  providing a robust validation of our Kerr--AdS calculation.\\

In this letter, we rely exclusively on the near-zone solution, avoiding the need for global asymptotic matching between the near-horizon and far-AdS regions. The usual matching techniques divide the spacetime into near and far zones, solving the field equations independently and joining them in an intermediate overlap region—a process that strictly constrains the domain to the low-frequency regime, $M\omega \ll 1$~\cite{stat1,c}. By extracting the frequency-dependent tidal response directly from the near-zone region, our framework offers a complementary and efficient alternative to global matching and effective field theory (EFT) approaches.\\

Our results strongly indicate that Kerr black holes in anti-de Sitter spacetime generically possess non-vanishing conservative and dissipative tidal responses for both dynamical and static cases under massless bosonic and fermionic perturbations. Since our present study focuses on the near-horizon approximation, it would be interesting to explore our computation within the world-line effective field theory framework \cite{eft5,eft6}, by investigating the scattering of massless bosonic and fermionic fields. Although our current study gives exact results within a background vacuum environment, the analysis can be extended to cases surrounded by matter fields. A related problem has been studied in Ref.\cite{rel}.

\end{document}